\documentstyle[12pt]{article}
\def\b{\begin{equation}}
\def\e{\end{equation}}

\begin{document}

\title{Tree-level Scattering Amplitude \\in de Sitter Space}
\author{
S. Rouhani $^{1}$ and M.V. Takook$^{1,2}$
\thanks{E-mail: takook@razi.ac.ir}}

\maketitle \centerline{\it $^{1}$Plasma Physics Research Centre,
Islamic Azad University,}
 \centerline{\it P.O.BOX 14835-157, Teheran,
IRAN} \centerline{\it  $^{2}$Department of Physics, Razi
University, Kermanshah, IRAN}
\vspace{15pt}

\begin{abstract}

In previous papers \cite{dere,gareta}, it was proved that a
covariant quantization of the minimally coupled scalar field in de
Sitter space is achieved through addition of the negative norm
states. This causal approach which eliminates the infrared
divergence, was generalized further to the calculation of the
graviton propagator in de Sitter space \cite{ta2} and one-loop
effective action for scalar field in a general curved space-time
\cite{ta5}. This method gives a natural renormalization of the
above problems. Pursuing this approach, in the present paper the
tree-level scattering amplitudes of the scalar field, with one
graviton exchange, has been calculated in de Sitter space. It is
shown that the infrared divergence disappears and the theory
automatically reaches a renormalized solution of the problem.

\end{abstract}

\vspace{0.5cm} {\it Proposed PACS numbers}: 04.62.+v, 98.80.Cq,
12.10.Dm
\newpage

\section{Introduction}

Positivity condition stemming from the concept of probability,
constitutes one of the principles of the axiomatic QFT
\cite{stwi}. Application of this principle {\it i.e.} the
positivity condition in the gauge QFT (GQFT), in Minkowski space,
results in breaking of the Lorentz covariance and the appearance
of the infrared divergence. It has been shown that a causal and
covariant GQFT \cite{st} can be obtained  through an indefinite
metric quantization ({\it i.e.} maintaining the negative norm
states: NNS). As a result the positivity condition must be
eliminated in GQFT in Minkowski space. The general properties of
this quantization , {\it i.e.} the indefinite metric Fock
quantization, were studied by Mintchev \cite{mi}.

The same problems of the GQFT appear in the quantization of the
free massless minimally coupled scalar field in de Sitter (dS)
space-time ({\it i.e.} appearance of the infrared divergence
\cite{ra} and the breaking of the dS covariance \cite{al}). It has
been shown that the presence of negative norm states are
indispensable for a fully covariant quantization of the minimally
coupled scalar field in de Sitter space \cite{dere,gareta}. This
type of NNS is different from of the NNS in the GQFT. So we call
this construction a ``Krein QFT'' (KQFT). This approach in the
free field quantization, while leaving the physical content of the
theory unchanged, removes the infrared and the ultraviolet
divergences. Following this pattern the normal ordering procedure
is rendered useless since the vacuum energy remains convergent
\cite{gareta}. The infrared divergence in the two-point function
disappears as well \cite{ta3}. The crucial element in the KQFT is
the presence of negative frequency solutions, which are
indispensable for preservation of the full covariance of the
problem. It has also been discussed that the application of this
method to the free ``massive'' scalar field in de Sitter space
results in an automatic and covariant renormalization of the
vacuum energy \cite{gareta}, as well.

The linear quantum gravity in dS space can be constructed from a
projector tensor and a massless minimally coupled scalar field
\cite{ta,gagarerota}. However, in so doing the above problems
appear again \cite{anilto1, altu}. It is also shown that the
infrared divergence or the pathological behavior of the graviton
propagator is gauge dependent and therefore should not appear in
an effective way as a physical quantity \cite{ta2,ta, vera,hahetu,
hiko1, hiko2}. An explicit construction of the covariant
quantization of the traceless rank-2 ``massless'' tensor field in
dS space (linear covariant quantum gravity) has been thoroughly
studied \cite{ta,gagarerota}. The main ingredient of this
construction is the presence of two different types of NNS. The
first one is similar to those non physical states, dual to gauge
states, which appear in GQFT in Minkowski space. The other is due
to the consideration of negative frequency solutions in order to
preserve de Sitter covariance. This is similar to the case of free
massless minimally coupled scalar field in dS space. This
construction allows us to avoid the infrared divergence, and
enables us to obtain a covariant two point function \cite{ta2}. It
should be noted that by adopting KQFT framework, various Green
functions for a scalar field, in a general curved space-time,
appear to be convergent in the ultraviolet limit. A natural
renormalized one-loop effective action has been also obtained
\cite{ta5}.

The appearance of the infrared divergence and the necessity of the
inclusion of the NNS for preservation of the covariance of the
problem constitutes the common denominator between GQFT in
Minkowski space and the calculation of massless minimally coupled
scalar field in dS space. In GQFT, however, a local symmetry is
present, where as in the  dS space, a global symmetry is the
central elements of the problem. The infrared divergence
disappears in the dS space case. In other words there are two
different types of NNS, one is due to the local gauge invariance
(sign of the metric) and the other is due to the global gauge
invariance (negative frequency solution). The important question
which deserves to be considered is: {\it Is it possible to
generalize the free KQFT in dS space, to the interaction field?}

By extending the KQFT method to the interacting quantum field in
Minkowski space-time ($\lambda\phi^4$ theory,) a natural
renormalization to the one-loop approximation has been achieved
\cite{ta4}. A very interesting consequence is that the Schwinger
commutator function, the retarded and the advanced Green
functions, are one and the same in both formalisms. Another
remarkable result is that the vacuum energy, automatically reduces
to zero \cite{ta4}.

In this paper the indefinite metric quantization {\it i.e.} the
presence of negative frequency states, as auxiliary unphysical
states, have been included for calculation of the tree-level
scattering amplitudes of the scalar field with one graviton
exchange in the de Sitter space. It is shown that the previous
divergence \cite{flilto}, disappears and the theory is
automatically renormalized.

\section{Tree-level scattering amplitudes}

Recalling the previous calculation of the tree-level scattering
amplitudes of two scalar particles in the one-graviton-exchange
approximation \cite{flilto}. An expression of the following form
must be calculated \b{\cal M}=\int
T^{\mu\nu}(x)G_{\mu\nu,\mu'\nu'}(x,x')T^{\mu'\nu'}(x')dV_xdV_{x'},\e
where $T^{\mu\nu}$ is a conserved, physical energy-momentum tensor
of the scalar field and $G_{\mu\nu,\mu'\nu'} $ is the graviton
two-point function. By explicit computation, it was shown that the
only growing part of ${\cal M}$, for large $z$, comes from the
transverse and traceless part of the graviton two-point function
\cite{flilto}, \b G_{TT}^{\mu\nu,\mu'\nu'}(x,x')\sim \ln (1-z)
{\cal D}^{\mu\nu,\mu'\nu'}\;\;\; \mbox{for large $z$},\e where
${\cal D}$ is a maximally symmetric bi-tensor.

For simplicity, a conformally coupled scalar field was considered
\b\phi(x)=\sum_k a_k f_k(t)e^{i\vec k. \vec x}+ C.C. ,\e where
$f_k(t)=\sqrt{\frac{2 |\vec k|}{4\pi^2}}t^2\frac{\sin t|\vec
k|}{t|\vec k|}$. A particular coordinate system is chosen \b
ds^2=-\frac{1}{t^2}dt^2+\frac{1}{t^2}d\vec x^2,\e in which
$z=1+\frac{1}{4tt'}[(t-t')^2-(\vec x-\vec x')^2]$. For the
S-channel diagram the only term, which depend on the scattering
angle $\vec k. \vec k'$, gives $$ {\cal
M}^{\theta\mbox{-dep.part}}\sim -8 (\vec k. \vec k')^2 \int
dV_xdV_{x'} t^2t'^2 f^2_k(t)f^2_{k'}(t')$$ \b \ln
(1-z)\left(1+\frac{4}{3}\frac{(\vec x-\vec x')^2}{(\vec x-\vec
x')^2- (t-t')^2}+\frac{8}{15}\left[\frac{(\vec x-\vec x')^2}{(\vec
x-\vec x')^2- (t-t')^2}\right]^2\right),\e where
$dV_x=\frac{1}{t^4}dtd^3x$. This term has a $\cos^2 \theta$
dependence characteristic of a spin-two exchange, but due to the
$\ln (1-z)$ element, it appears to be divergent.

At this stage, the new method of field quantization is
implemented. In this method the field operator $(3)$ is defined by
\cite{gareta} \b\phi(x)=\frac{1}{\sqrt 2}\sum_k a_k f_k(t)e^{i\vec
k. \vec x}+ b_k f_k^*(t)e^{-i\vec k. \vec x}+C.C. ,\e where $$
[a(\vec k),a^{\dag}(\vec k')]=\delta(\vec k-\vec k')
,\;\;a^{\dag}(\vec k)\mid 0>= \mid 1_{\vec k}>=\mid
\mbox{one-particle state} >,$$ $$ [b(\vec k),b^{\dag}(\vec
k')]=-\delta(\vec k-\vec k') ,;\;\;b^{\dag}(\vec k)\mid 0>=  \mid
\bar1_{\vec k}>=\mid \mbox{unphysical state} >$$ $$ b(\vec k)\mid
\mbox{physical state}
>=0;\;\; a(\vec k)\mid \mbox{unphysical state} >=0,$$ \b <
\mbox{unphysical state}\mid \mbox{physical state} >=0.\e By using
the above relations it can be shown that the stress tensor for the
physical state is not affected by the negative norm state
(\cite{gareta} section 6). Therefore, the only departure from
previous method appears in the graviton two-point function. In
this case the logarithmic term ($\ln (1-z)$) must be replaced by
\cite{ta2,ta3,gagarerota} \b \mbox{const.} \;\epsilon
(X^0-X'^0)\theta (z-1),\e where \b X^0=\sinh
t+\frac{1}{2}e^{t}|\vec x|^2 ,\;\;\;\epsilon
(X^0-X'^0)=\left\{\begin{array}{clcr} 1&X^0>X'^0 \\ 0&X^0=X'^0\\
-1&X^0<X'^0,\\    \end{array} \right. \e and $\theta$ is the
Heaviside step function.  This expression (eq.($8)$) can also be
obtained by the use of the fact that the Krein-Feynman propagator
is the real part of the previous Feynman propagator (eq.($13a)$ of
\cite{tswo}.) By the application of the equation $(8)$ in the
integral $(5)$, we obtain
$$ {\cal M}^{\theta\mbox{-dep.part}}\sim \mbox{const.}\; (\vec k.
\vec k')^2 ,$$ which is free of any infrared divergence.

By the very same procedure, it can also be shown that the
effective action, in the one-loop approximation, for the
large-distance behavior, is convergent. This result reaffirms the
Iliopoulos's calculation \cite{anilto1}.

\section{Conclusion}

The appearance of singularities in physical quantities in QFT, is
a manifestation of presence of anomalies in QFT. For eliminating
these anomalies, the normal ordering procedure and the
renormaization techniques are applied to QFT.

The covariant quantization of the free minimally coupled scalar
field in de Sitter space in absence of the positivity condition,
and its consequences has been studied \cite{gareta}. In this case
without changing the physical content of the theory, an
automatically renormalized results was obtained. Following this
procedure, the covariant principle in dS space has been conserved
as well. In the interaction case, the tree-level scattering
amplitudes of the scalar field, with one graviton exchange in de
Sitter space, has been calculated by this method. It is shown that
the previous infrared divergence disappears, and the theory is
automatically renormalized as well.

\vskip 0.5 cm \noindent {\bf{Acknowledgements}}: The authors would
like to thank M. Tanhay for his interest in this work.

\end{document}